\documentclass[pra,aps,superscriptaddress,amsmath,amssymb,preprint]{revtex4}
\usepackage{CJKutf8}
\usepackage{graphicx}
\usepackage{dcolumn}
\usepackage{bm,braket}

\usepackage{mathptmx}
\usepackage{graphicx}
\usepackage{subfigure}
\usepackage{amsmath}
\usepackage{amsfonts}
\usepackage{amssymb}
\usepackage{mathrsfs}
\usepackage{subfigure}
\usepackage{xcolor}
\usepackage[colorlinks=true,linkcolor=blue,anchorcolor=red,citecolor=blue, urlcolor=blue]{hyperref}
\usepackage{ulem}

\makeatletter
\renewcommand*\env@matrix[1][\arraystretch]{%
  \edef\arraystretch{#1}%
  \hskip -\arraycolsep
  \let\@ifnextchar\new@ifnextchar
  \array{*\c@MaxMatrixCols c}}
\makeatother

\def\be{\begin{equation}}
\def\ee{\end{equation}}
\def\ba{\begin{eqnarray}}
\def\ea{\end{eqnarray}}

\newlength{\seplinewidth}
\newlength{\seplinesep}
\colorlet{sepline}{orange}

\begin{document}
\begin{CJK*}{UTF8}{gbsn}

\title{Quantum Computing by Cooling}

\author{Jiajin Feng(冯嘉进)}

\affiliation{International Center for Quantum Materials, School of Physics, Peking University, Beijing 100871, China}

\author{Biao Wu(吴飙)}
\email{wubiao@pku.edu.cn}
\affiliation{International Center for Quantum Materials, School of Physics, Peking University, Beijing 100871, China}
\affiliation{Wilczek Quantum Center, School of Physics and Astronomy, Shanghai Jiao Tong University, Shanghai 200240, China}
\affiliation{Collaborative Innovation Center of Quantum Matter, Beijing 100871, China}

\author{Frank Wilczek}
\affiliation{Center for Theoretical Physics, MIT, Cambridge, Massachusetts 02139, USA}
\affiliation{T. D. Lee Institute, Shanghai Jiao Tong University, Shanghai 200240, China}
\affiliation{Wilczek Quantum Center, School of Physics and Astronomy, Shanghai Jiao Tong University, Shanghai 200240, China}
\affiliation{Department of Physics, Stockholm University, Stockholm SE-106 91, Sweden}
\affiliation{Department of Physics and Origins Project, Arizona State University, Tempe, Arizona 25287, USA}

\date{\today}

\begin{abstract}
Interesting problems in quantum computation take the form of finding low-energy states of (pseudo)spin systems
with engineered Hamiltonians that encode the problem data.  Motivated by the practical possibility of producing
very low-temperature spin systems, we propose and exemplify the possibility to compute by coupling the
computational spins to a non-Markovian bath of spins that serve as a heat sink.  We demonstrate both analytically and numerically
that this strategy can achieve quantum advantage in the Grover search problem.
\end{abstract}
\preprint{MIT-CTP/5300}
\maketitle
\end{CJK*}

\section{Introduction}

Quantum computing can be implemented, conceptually, using either quantum logic gates ~\cite{Grover1996,Shor1999,Takeshita2020,Hendrickx2020,Petit2020} or Hamiltonians ~\cite{Farhi1998,Farhi2000}. Under broad assumptions the two techniques are computationally equivalent, abstractly ~\cite{Dam2007,YHW}, but each brings in different intuitions. Roughly speaking, the gate approach is more familiar in the analysis of Turing machines and practical digital circuits, while a Hamiltonian approach is more familiar in the analysis of natural physical systems.  The quantum adiabatic approach to optimization problems ~\cite{Farhi2000,vanDam2001,Ozfidan2020} is an outstanding example of a class of algorithms suggested by a physical phenomenon, i.e., the  preservation of quantum ground states under adiabatic evolution; other examples include algorithms inspired by resonance \cite{Wilczek2020} and diffusion \cite{QTree}.   Physics can also suggest possibilities for resources that are not usually considered in the standard conceptual models, e.g. global addressing of qubits by external fields or controlled coupling to physically realistic heat sinks, as exemplified below.

The observation that many important computational problems can be encoded as the search for low-energy states of explicit, deceptively simple Hamiltonians $H_{\rm s}$ is central to applications of the adiabatic algorithm.  One way to bring a system to low energy, of course, is to couple it to low temperature system.  The production of (pseudo)spin systems with very low temperature is a highly developed art \cite{Valenzuela2006,XuXiaodong2007,Press2008,Togan2011,Yang2020}.  Putting those observations together, we are led to consider the possibility of addressing computational problems by coupling systems whose ground states contain the answer- ``computational qubits'' -  to systems that have very low temperatures - ``bath qubits'' -  that act as an energy sink.

The issue then arises, whether this procedure can be performed in a way that maintains an advantage of quantum over classical computation.
Here we demonstrate that it can, at least in the context of the iconic Grover search problem \cite{Grover1996,Grover1997,Grover1998,Nielsen2010}.


We propose a general quantum cooling algorithm to find the ground state of a problem Hamiltonian  $H_{\rm s}$.
The  problem system is coupled to a non-Markovian quantum bath, which is chosen to be an interacting (pseudo)spin system
with trivial and easy-to-prepare ground states. As a result, the quantum
bath  can be readily set to the ground state.  Because the bath is effectively at zero temperature, the energy
will flow from the system into the bath and the problem system is cooled down to its ground state.
The cooling speed of our algorithm is affected  by various factors, such as
the effective interaction between the system and the bath, and their energy gaps.

To show that our cooling algorithm incorporates essentially quantum features, different from classical thermal cooling \cite{vanDam2001,Edward2002},
we set up two different cooling algorithms to do random search. In the first algorithm, the coupling between the system and the bath
is simple but non-local. The analytical solution shows that its time complexity is $O\left( \sqrt{N_{\rm s}}\right)$ ($N_{\rm s}$ is the
dimension of the Hilbert space of $H_{\rm s}$). In the second algorithm,
the coupling is local. Our analysis and numerical computation find that  the time
complexity  is $\sim O(N_{\rm s}^{0.55})$. Both of the algorithms are  faster than the classical time complexity  $O(N_{\rm s})$,
showing our cooling scheme is quantum coherent and different from cooling with a Markovian bath.

\section{Cooling with Quantum Bath}
\subsection{General framework}
Our computing scheme involves two separate sets of qubits: computational qubits and bath qubits, for which the problem Hamiltonian
$H_{\rm s}$ and the bath Hamiltonian $H_{\rm b}$ are constructed, respectively.
The problem Hamiltonian  $H_{\rm s}$ encodes the solutions of a given problem in  its ground states. The bath Hamiltonian $H_{\rm b}$ is usually
an interacting spin system with trivial ground states, so that it can be brought close to absolute zero temperature readily. For example, one may choose
\be
H_{\rm b}=-J \sum_{\braket{m,m'}}\left(\hat{\sigma}^x_m\hat{\sigma}^x_{m'}+\hat{\sigma}^y_m\hat{\sigma}^y_{m'}+\hat{\sigma}^z_m\hat{\sigma}^z_{m'}\right) \, ,
\label{eq:XXX}
\ee
where $J>0$ and $\hat{\sigma}^{x,y,z}_m$ is the Pauli matrix of the $m$th spin. The summation is over an arbitrary set of qubit pairs $\braket{m,m'}$. This Hamiltonian has at least two trivial ground states $\ket{00\cdots 0}$ and $\ket{11\cdots 1}$ ($\ket{0}$ for spin-down
and $\ket{1}$ for spin-up), which are easy to be prepared. When the spins sit on a one-dimensional chain with the nearest neighbor interaction,
it is the well-known Heisenberg XXX model \cite{Fabio2017,Gromov2017,Salberger2017}, and its spin wave excitation can carry energy away from the problem system \cite{Jepsen2020,Bertini2016,Castro2016}.  There are many
interacting spin systems with trivial ground states~\cite{Yanglin}.

The total Hamiltonian for our cooling algorithm is
\be
H=H_{\rm s}+H_{\rm b}+H_{\rm I}\,,
\ee
where $H_{\rm I}$ is the coupling between computational qubits and bath qubits. If there are $n_{\rm s}$ computational qubits and
$n_{\rm b}$ bath qubits, the Hilbert space size  is $N_{\rm s}=2^{n_{\rm s}}$ for $H_{\rm s}$ and
 is $N_{\rm b}=2^{n_{\rm b}}$ for $H_{\rm b}$.  Their energy eigen-equations are
$H_{\rm s}\ket{\psi_{i_{\rm s}}}=E_{i_{\rm s}}\ket{\psi_{i_{\rm s}}}$ and $H_{\rm b}\ket{\phi_{j_{\rm b}}}=E_{j_{\rm b}}\ket{\phi_{j_{\rm b}}}$,
respectively. The total Hilbert space of size $N_{\rm c}=N_{\rm s}N_{\rm b}$ is spanned by the base
$\ket{\psi_{i_{\rm s}}}\otimes\ket{\phi_{j_{\rm b}}}\equiv\ket{\psi_{i_{\rm s}},\phi_{j_{\rm b}}}$.
Among all $\ket{\psi_{i_{\rm s}}}$'s and $\ket{\phi_{j_{\rm b}}}$'s, for clarity, we use $\ket{g_{\rm s}}$ to denote
the unknown ground states of the problem system $H_{\rm s}$ which are the solutions of the problem,
and  $\ket{g_{\rm b}}$  the known ground state of the bath $H_{\rm b}$ which is easy to be prepared. We set $\protect \hbar =1$
and consider $E$ and $t$ as dimensionless variables in the following discussion because they are irrelevant to time complexity,
which is our focus.

We intend to use the bath to cool down the problem system and find its ground states $\ket{g_{\rm s}}$.
The bath is initialized in one of its trivial ground states, so that it is at the absolute zero temperature.
The problem system can be initialized in an arbitrary state that is easy to be prepared.
So, the full initial wave function at $t=0$ is
\begin{eqnarray}
\ket{\Psi_{\rm in}}=\sum_{i_{\rm s}=0}^{N_{\rm s}-1} c_{i_{\rm s}} \ket{\psi_{i_{\rm s}},g_{\rm b}}   \ ,
\label{eq:Psi0}
\end{eqnarray}
where $c_{i_{\rm s}}$ is the superposition probability amplitude. Once the interaction $H_{\rm I}$ is turned on,
the whole composite system starts evolution with $\ket{\Psi}=e^{-iHt}\ket{\Psi_{\rm in}}$ and
the energy will flow from the problem system to the bath. As a result, the problem system is cooled and will get closer
to its ground state. If we measure the problem system at the end of cooling, we will have the following probability for
finding the ground state $\ket{g_{\rm s}}$ of the problem system $H_{\rm s}$,
\begin{eqnarray}
P_g &=& \sum_{j_{\rm b}=0}^{N_{\rm b}-1} \left| \langle g_{\rm s},\phi_{j_{\rm b}}| \Psi\rangle \right|^2 \ .
\label{eq:psg}
\end{eqnarray}
The aim of our cooling algorithm is to make this probability high in a shortest time.

Here are key features of our cooling scheme.
\begin{itemize}
\item It is different from cooling with a Markovian thermal bath.  All the processes here are quantum coherent.
\item As the bath has easy-to-prepare ground states, it can be reset to zero temperature whenever it is necessary.
\item Large density of states of the bath is required.
Because efficient quantum transitions occur at energies corresponding to the spacing of computational levels which the final state $\ket{\psi_{i_{\rm s}'},\phi_{j'_{\rm b}}}$
have similar energy to the initial state $\ket{\psi_{i_{\rm s}},g_{\rm b}}$,
namely, $E_{i_{\rm s}'}+E_{j'_{\rm b}}\approx E_{i_{\rm s}}+E_{g_{\rm b}}$.  Note that in many important optimization problems the eigen-energies of $H_{\rm s}$ are integer multiples of a single parameter $\Delta$.
\item The number of states in the bath should  increase rapidly with energy. This encourages the bath to occupy higher energy states
and absorb energy from the problem system.
This is  satisfied in most many-body systems, where higher energy can excite more quasi-particles. If the quasi-particles are weakly interacting, the growth is exponential.
\item The total Hamiltonian is unchanged during the evolution.
This helps maintain quantum coherence.
\end{itemize}

Our quantum cooling algorithm differs from the heat-bath algorithmic cooling (HBAC), quantum-circuit refrigerator (QCR) and interaction enhanced
quantum computing.
HBAC is used to purify a known ground state \cite{Boykin2002,RB2016,Raeisi2019,Sadegh2,Zaiser2021}. QCR is an open system usually coupled to Markovian bath \cite{Tan2017,Silveri2017,Hsu2020}. For interaction enhanced
quantum computing, the interaction is between different quantum computers not between a system and a bath~\cite{Wenxian}.
We also note that an early work indicates that non-Markovian bath could improve the performance of a quantum refrigerator~\cite{Camati2020}.

\subsection{Toy Model}

\begin{figure}[b]
\begin{center}
\includegraphics[clip = true, width =0.9\columnwidth]{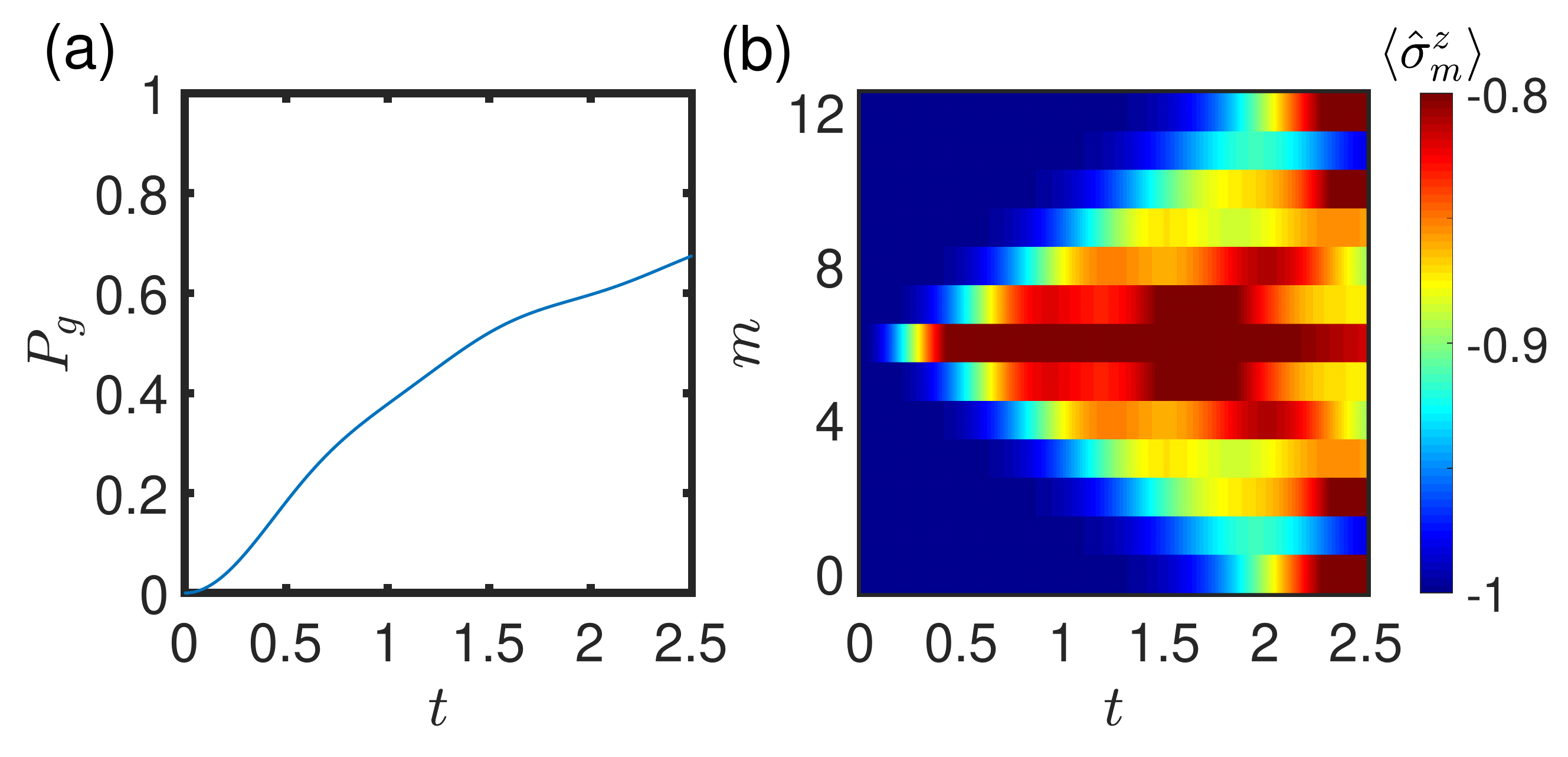}
\caption{\label{fig:wave}(color online) (a) The ground state probability of the system in Eq. (\ref{eq:Hone}).
(b) The color is the $z$ direction component of each qubit in the bath.  $m$ marks different qubits. Other parameters are $n_{\rm s}=1$, $n_{\rm b}=13$, $J=1$, $B=1$, $\lambda=1$ and $\ket{\Psi_{\rm in}}=\ket{e_{\rm s},g_{\rm b}}$. }
\end{center}
\end{figure}

To get oriented, let us briefly consider a toy example.
The system is a single spin coupled to the middle spin of a one-dimensional spin chain,
\begin{eqnarray}
H_{\rm s} = B\hat{s}^z \ , \quad H_{\rm I} = \lambda\hat{s}^y\hat{\sigma}_{\left\lfloor \frac{n_{\rm b}}{2}\right\rfloor}^y \ ,
\label{eq:Hone}
\end{eqnarray}
where $\hat{s}^z_m$ is the Pauli matrix of the system, $B$ is the on-site energy and $\lambda$ is the coupling strength. The bath is one-dimensional spin chain governed by the Hamiltonian
in Eq. (\ref{eq:XXX}) with the nearest neighbor interaction and periodic boundary condition.

The system spin is set in the excited state and the bath is set in the ground state with all spins down.
After the interaction is turned on instantaneously,  the energy begins to flow into the bath, generating spin wave excitations that
carry away energy from the problem system \cite{VANDAELE201739,Chuanpu2018,Bertini2016,Castro2016}. Numerical results are shown in Fig. \ref{fig:wave}.
In Fig. \ref{fig:wave}(a), the probability of the system in the ground state becomes larger with time. Meanwhile, the energy spreads away from the middle of the chain as shown in Fig. \ref{fig:wave}(b) [see Appendix \ref{app:wave} for an analytical approach].

\section{Unsorted Search}

Unsorted search is a benchmark example demonstrating a sharp difference between quantum and classical computers.
To search $M$ targets among $N$ unsorted items, the time complexity of a classical algorithm is $O(N/M)$. In contrast,
the Grover's algorithm on a quantum computer has  time complexity of  $O\left( \sqrt{N/M}\right)$ \cite{Boyer1998,SearchRev2017}.
When our cooling algorithm is applied to this search problem, we expect a time complexity no better than
$O\left( \sqrt{N_{\rm c}/N_{\rm b}}=\sqrt{N_{\rm s}}\right)$. The reason is that all the $N_{\rm b}$ states $\ket{g_{\rm s},j_{\rm b}}$'s are
the targets among the total $N_{\rm c}=N_{\rm s}N_{\rm b}$ states for the whole system.
We present two different cooling algorithms for unsorted search: one with non-local interaction and the other with local interaction.
The first achieves the benchmark quantum time complexity $O\left(\sqrt{N_{\rm s}}\right)$ and the second comes close to that time complexity
$\sim O\left(N_{\rm s}^{0.55}\right)$.

\begin{figure}[!tb]
\begin{center}
\includegraphics[clip = true, width =0.8\columnwidth]{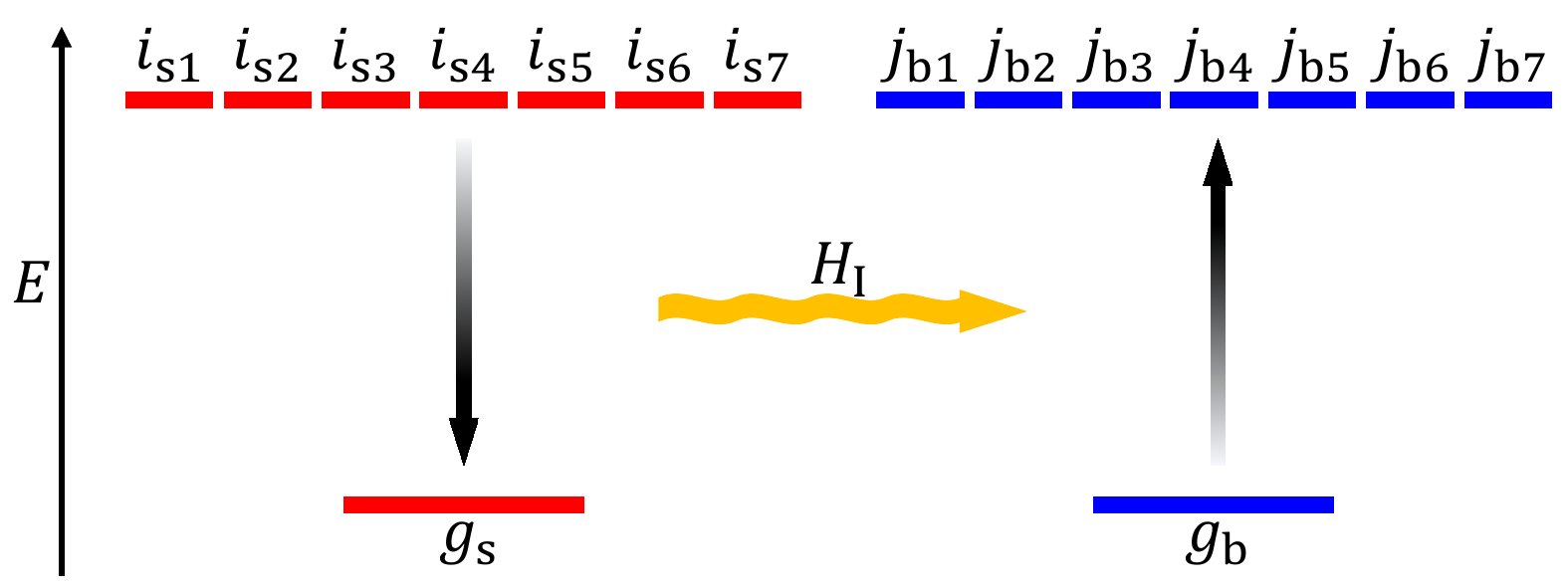}
\caption{\label{fig:setup} (color online) The diagram of the Hamiltonian $H=H_{\rm s}+H_{\rm b}+H_{\rm I}$ for unsorted search.
The red bars are the energy levels of $H_{\rm s}$. The blue bars are the energy levels of $H_{\rm b}$.}
\end{center}
\end{figure}

In our quantum algorithm, all the search items are stored in system qubits and represented by states $\ket{i_{\rm s}}$.  In the state
of $\ket{i_{\rm s}}$,  the $m$th system qubit  is in the state $\ket{i_{\rm s}^{(m)}}$ ($m=0,1,2,\cdots,n_{\rm s}-1$) with
$i_{\rm s}^{(m)}$ being the binary digit of $i_{\rm s}$. For simplicity, we consider the case where there is only one target, $\ket{x_{\rm s}}$,
which is one of the $\ket{i_{\rm s}}$'s. We construct two Hamiltonians, respectively, for the problem system
and  the bath as \cite{Farhi1998,vanDam2001,Cerf2000,Wilczek2020}
\begin{equation}
H_{\rm s}=-\ket{g_{\rm s}}\bra{ g_{\rm s}}~\,,~~~~~
H_{\rm b}=-\ket{g_{\rm b}} \bra{g_{\rm b}} \ .
\label{eq:H0}
\end{equation}
These two Hamiltonians have only two eigen-energies respectively, one for non-degenerate ground state
and the other for highly-degenerate excited states (see Fig. \ref{fig:setup}).
It is important to note that the system ground state $\ket{g_{\rm s}}=\ket{x_{\rm s}}$ is unknown while
the bath ground state $\ket{g_{\rm b}}$ is known and can be assumed to be $\ket{g_{\rm b}}=\ket{000\cdots 0}$ without loss
of generality. For the above two Hamiltonians, their energy-eigenstates are  $\ket{\psi_{i_{\rm s}}}=\ket{i_{\rm s}}$ and
$\ket{\phi_{j_{\rm b}}}=\ket{j_{\rm b}}$, respectively.

Our quantum algorithm is to find the system's ground state $\ket{g_{\rm s}}$ by coupling the system to the bath and
taking advantages that the bath ground state $\ket{g_{\rm b}}$ is known
and easy to be prepared. Below are two quantum algorithms with different couplings, both of which outperform the classical algorithm.

\subsection{Non-local Interaction}
Here we choose the following non-local  interaction to couple the system to the bath,
\begin{eqnarray}
H_{\rm I} &=&-\ket{\xi}\bra{\xi} \ ,
\label{eq:Hinl}
\end{eqnarray}
where $\ket{\xi}=\sqrt{1/N_{\rm c}}\sum_{i_{\rm s}=0}^{N_{\rm s}-1}\sum_{j_{\rm b}=0}^{N_{\rm b}-1} \ket{i_{\rm s},j_{\rm b}}$.
Similar non-local interactions can be found in Ref. \cite{Farhi1998,vanDam2001,Cerf2000,Wilczek2020} and their justification
can be found in  Appendix \ref{app:oracle}. The initial state for the whole system is
\begin{equation}
 \ket{\Psi_{\rm in}}=\frac{1}{\sqrt{N_{\rm s}}}\sum_{i_{\rm s}=0}^{N_{\rm s}-1} \ket{i_{\rm s},g_{\rm b}}\,,
 \label{eq:sortin}
\end{equation}
 where the bath is in the ground state.
 Once the interaction is turned on, energy will flow from the problem system to the bath and the problem system will be cooled down to $\ket{ g_{\rm s}}$.

For this special case,  the whole cooling process is confined in a subspace spanned by the following four  states,
\begin{eqnarray}
\ket{Y}&=&\frac{1}{\sqrt{N_{\rm c}-N_{\rm s}-N_{\rm b}+1}}\sum_{i_{\rm s}=0,\atop i_{\rm s}\neq g_{\rm s}}^{N_{\rm s}-1}\sum_{j_{\rm b}=0,\atop j_{\rm b}\neq g_{\rm b}}^{N_{\rm b}-1}\ket{i_{\rm s},j_{\rm b}} \,,\label{eq:Y}\\
\ket{\beta}&=&\frac{1}{\sqrt{N_{\rm b}-1}}\sum_{j_{\rm b}=0,\atop j_{\rm b}\neq g_{\rm b}}^{N_{\rm b}-1} \ket{g_{\rm s},j_{\rm b}}\,, \label{eq:beta}\\
\ket{\alpha}&=&\frac{1}{\sqrt{N_{\rm s}-1}}\sum_{i_{\rm s}=0,\atop i_{\rm s}\neq g_{\rm s}}^{N_{\rm s}-1} \ket{i_{\rm s},g_{\rm b}}\,,\label{eq:alpha} \\
\ket{G}&=&\ket{g_{\rm s},g_{\rm b}}\,. \label{eq:G}
\end{eqnarray}
In other words, the Hamiltonian is effectively a $4\times4$ matrix [see Appendix \ref{app:H44}].
For brevity, we just present the Hamiltonian in the limit of $1\ll N_{\rm s},N_{\rm b}\ll N_{\rm c}$
\begin{eqnarray}
H &\approx&-\ket{Y}\bra{Y}-\ket{\alpha}\bra{\alpha}-\ket{\beta}\bra{\beta}-2\ket{G}\bra{G} \nonumber\\
&&-\sqrt{\frac{N_{\rm s}}{N_{\rm c}}}\left( \ket{Y}\bra{\alpha}+\ket{\alpha}\bra{Y} \right) \nonumber\\
&&-\sqrt{\frac{N_{\rm b}}{N_{\rm c}}}\left( \ket{Y}\bra{\beta}+\ket{\beta}\bra{Y} \right) \ .
\label{eq:simH44}
\end{eqnarray}
This matrix can be diagonalized exactly.  As $\ket{\Psi_{\rm in}}=\sqrt{1/N_{\rm s}}\ket{G}+\sqrt{(N_{\rm s}-1)/N_{\rm s}}\ket{\alpha}$,
its time evolution is
\begin{eqnarray}
 && \ket{\Psi} \approx e^{-2it}\sqrt{\frac{1}{N_{\rm s}}}\ket{G}+e^{-it}\sqrt{\frac{N_{\rm s}-1}{N_{\rm s}}}\left[ \frac{N_{\rm s}\cos\omega t+N_{\rm b}}{N_{\rm s}+N_{\rm b}}\ket{\alpha} \right. \nonumber\\
  &&\left. +\frac{\sqrt{N_{\rm s}N_{\rm b}}(\cos\omega t-1)}{N_{\rm s}+N_{\rm b}}\ket{\beta}+ i\sqrt{\frac{N_{\rm s}}{N_{\rm s}+N_{\rm b}}}\sin\omega t\ket{Y} \right] \,,
\label{eq:Psit}
\end{eqnarray}
where the oscillation  frequency is
\begin{eqnarray}
\omega &\approx& \sqrt{\frac{N_{\rm s}+N_{\rm b}}{N_{\rm c}}} \ .
\end{eqnarray}

 We can substitute Eq. (\ref{eq:Psit}) into Eq. (\ref{eq:psg}) and get
\begin{eqnarray}
P_g &\approx& \frac{4N_{\rm s}N_{\rm b}}{(N_{\rm s}+N_{\rm b})^2}\sin^4\frac{\omega t}{2} \ .
\label{eq:pb}
\end{eqnarray}
For the special case $N_{\rm b}=N_{\rm s}$, we have $P_g\approx 1$ at $t=\pi\sqrt{N_{\rm s}/2}$. The time complexity of
our algorithm is $O(\sqrt{N_{\rm s}})$ that is as good as Grover's~\cite{Grover1996}. In general,
the average time needed to finish this algorithms is
\begin{eqnarray}
\overline{T}= \frac{\pi}{\max\left(P_g\right)_t\omega} = \frac{\pi\left(N_{\rm s}+N_{\rm b}\right)^{1.5}}{4\sqrt{N_{\rm s}N_{\rm b}}} \ .
\end{eqnarray}
When $N_{\rm b}=0.5N_{\rm s}$, the required time is shortest with $\overline{T}=2.04\sqrt{N_{\rm s}}$.
When $N_{\rm b}\ll N_{\rm s}$, the time complexity is $O(N_{\rm s})$, which is similar to the classical algorithm.
The reason is that there are not enough high energy states in a small bath to absorb energy. When $N_{\rm b}\gg N_{\rm s}$,
the time complexity is $O(N_{\rm b}/\sqrt{N_{\rm s}})$ because the effective interaction becomes small.
These results show that by  choosing the Hamiltonians properly we can  get the ground state of problem system efficiently
by coupling to a quantum bath.

\subsection{Local interaction}
\begin{figure}[!tb]
\begin{center}
\includegraphics[clip = true, width =0.9\columnwidth]{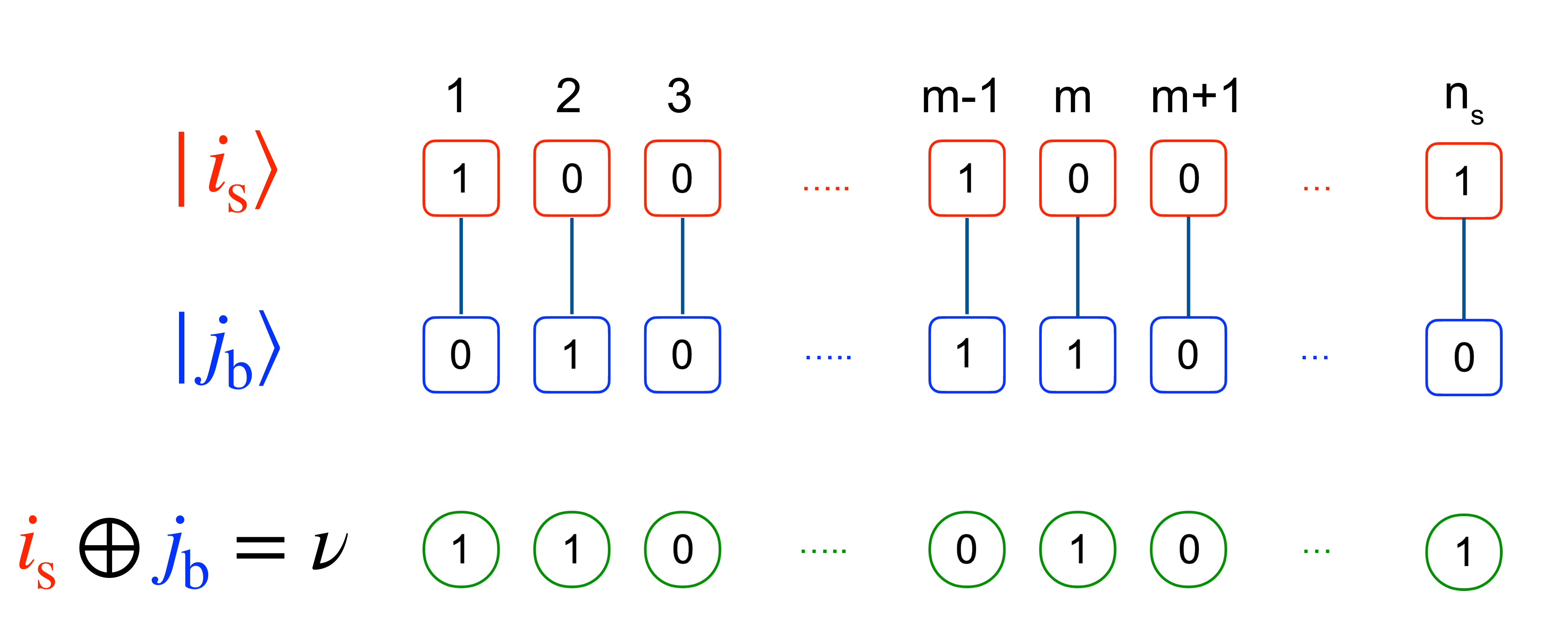}
\caption{\label{fig:chain} (color online) Illustration of $i_{\rm s}\oplus j_{\rm b}=\nu$. Red squares represent qubits of the system;
blue squares represent qubits of the bath; the verticle lines represent the pair-wise interaction $\hat{s}_m^x\hat{\sigma}_m^x$
between the system qubits and bath qubits.  The binary digits of $\nu$ are placed in circles for clarity.   }
\end{center}
\end{figure}

Our cooling algorithm can also achieve speed-up over the classical algorithm with local interactions.
We focus on the case where the number of bath qubits $n_{\rm b}$ is the same as the computational qubits $n_{\rm s}$, i.e.,
$n_{\rm b}=n_{\rm s}$.  The local interaction  is
\begin{eqnarray}
H_{\rm I}=-\lambda_{n_{\rm s}}\sum_{m=0}^{n_{\rm s}-1}\hat{s}_m^x\hat{\sigma}_m^x \ ,
\label{eq:Hil}
\end{eqnarray}
where $\hat{s}^x_m$ and $\hat{\sigma}^x_m$ acts on the $m$th qubit of the problem system and the bath, respectively.
$\lambda_{n_{\rm s}}$ is the interaction strength that $\lim_{n_{\rm s}\rightarrow\infty}\lambda_{n_{\rm s}}\times n_{\rm s}$ is a constant.
It makes $\langle H_{\rm I} \rangle$ and $\langle H_{\rm s} \rangle$ the same order of magnitude. This composite system can be viewed
as two parallel spin chains with pair-wise coupling (see Fig. \ref{fig:chain}).

\begin{figure}[tb]
\begin{center}
\includegraphics[clip = true, width =0.9\columnwidth]{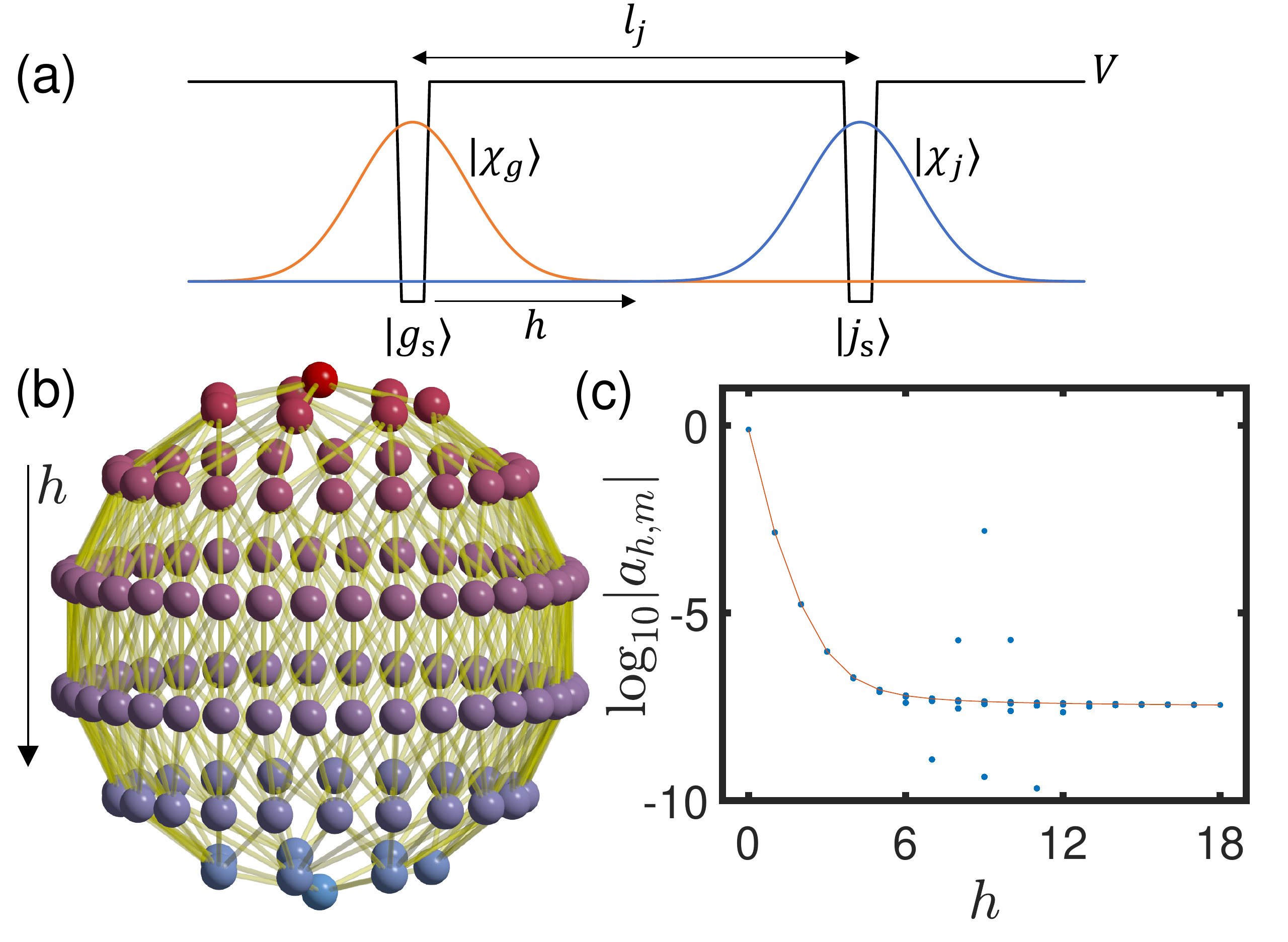}
\caption{\label{fig:decay} (color online) (a) One dimensional  schematic illustration of Hamiltonian $H_{j_{\rm s}}$. $\ket{\chi_g}$ and $\ket{\chi_j}$
are the low energy wave packets in the wells.
The combinations $(\ket{\chi_g}\pm \ket{\chi_j})/\sqrt{2}$ are the familiar ground state and the first excited state in the double-well.
(b) Schematic of the $n_{\rm s}$-dimensional hypercube (placed on a hypersphere). Each point represents one configuration of qubits
with the top point representing  $\ket{g_{\rm s}}$. The color represents the Hamming distance $h$ of $\ket{i_{\rm s}}$ from $\ket{g_{\rm s}}$
(red for smaller distance, blue for larger distance). Yellow lines represent hopping between different $\ket{i_{\rm s}}$.
(c) The actual wave function of one wave packet $\ket{\chi_g}$ (blue dots) reconstructed by diagonalizing Eq. (\ref{eq:Hj2}).
$\ket{j_{\rm s}}$ locates at $l_j=9$ for example. The number of qubits is $n_{\rm s}=18$.
The interaction strength is $\gamma_1=1$, $\gamma_2=1.16$ and $\gamma_{m\geqslant3}=0$. The calculated well component is $|a_0|^2=0.8$.
The orange line is the median among same $h$. }
\end{center}
\end{figure}

The dynamics governed by $H$ is a unitary evolution in a Hilbert space of dimension $N_{\rm c}=N_{\rm s}N_{\rm b}=N_{\rm s}^2$. Fortunately,
it can be decomposed into $N_{\rm s}$ independent dynamics with each of them restricted in a $N_{\rm s}$-dimensional Hilbert space.
The dynamics in each of these $N_{\rm s}$-dimensional Hilbert spaces is effectively a double-well tunneling  in an $n_{\rm s}$-dimensional
hypercube (see Fig. \ref{fig:decay}).

This decomposition is possible due to a special property of this system, which we call parity between system qubits
and bath qubits.  For a pair of states $\ket{i_{\rm s}}$ and $\ket{j_{\rm b}}$,
this parity is given by a number $\nu=i_{\rm s}\oplus j_{\rm b}$,  where $\oplus$ is a bitwise
module 2 addition as illustrated in Fig. \ref{fig:chain} (see Appendix \ref{app:xor} for more details).
Since $[\hat{s}_m^z\hat{\sigma}_m^z,H]=0$, the parity number $\nu$ is conserved during the dynamical evolution.

We define a sub-Hilbert space ${\mathcal H}_{\nu}$, which is spanned
by all  $\ket{i_{\rm s}, j_{\rm b}}$'s satisfying  $i_{\rm s}\oplus j_{\rm b}=v$. It is easy to check
that $j_{\rm b}=i_{\rm s}\oplus v$ if $v=i_{\rm s}\oplus j_{\rm b}$.
This means that in each subspace ${\mathcal H}_{\nu}$,
there is one to one mapping between the  system states $\ket{i_{\rm s}}$ and the bath states $\ket{j_{\rm b}}$.
Therefore, each Hilbert space  ${\mathcal H}_{\nu}$ is of dimension  $N_{\rm s}$.
The subspace ${\mathcal H}_{\nu}$ is invariant under the unitary transformation of the total Hamiltonian $H$.
As a result, the whole dynamical evolution is just a simple summation of dynamics in each subspace  ${\mathcal H}_{\nu}$.

We still choose Eq. (\ref{eq:sortin}) as the initial state, where different $i_{\rm s}$'s belong to different subspaces
${\mathcal H}_{v_{i_{\rm s}}}$ labelled by $v_{i_{\rm s}}=i_{\rm s}\oplus g_{\rm b}$.
Therefore, we can independently investigate the dynamical evolution within each subspace.
In a given subspace ${\mathcal H}_{v_{j_{\rm s}}}$ ($j_{\rm s}$ is one of $i_{\rm s}$'s),
there are only two on-site energy terms in Eq. (\ref{eq:H0}) and the total Hamiltonian is reduced to
\begin{eqnarray}
H_{j_{\rm s}}=-\ket{ g_{\rm s},j_{\rm b}}\bra{ g_{\rm s},j_{\rm b} }-\ket{j_{\rm s},g_{\rm b}}\bra{j_{\rm s},g_{\rm b}}-\lambda_{n_{\rm s}}
\sum_{m=0}^{n_{\rm s}-1}\hat{s}_m^x\hat{\sigma}_m^x \ . \nonumber\\
\label{eq:Hj}
\end{eqnarray}
In the subspace ${\mathcal H}_{v_{j_{\rm s}}}$, there is one-to-one mapping between $\ket{i_{\rm s}}$ and $\ket{j_{\rm b}}$
via $i_{\rm s}\oplus  j_{\rm b}=v_{j_{\rm s}}$.  As a result, we can hide the bath qubits and simplify
 the above Hamiltonian in the subspace as
\begin{eqnarray}
H_{j_{\rm s}}&=&-\ket{ g_{\rm s}}\bra{ g_{\rm s}}-\ket{j_{\rm s}}\bra{ j_{\rm s}}-\lambda_{n_{\rm s}}\sum_{m=0}^{n_{\rm s}-1}\hat{s}_m^x\,.
\label{eq:Hj2}
\end{eqnarray}
The system described by this Hamiltonian can be visualized as a particle living on a hypercube of $n_{\rm s}$ dimensions (see Fig. \ref{fig:decay}(b)).
Each site of this hypercube is represented by a state $\ket{i_{\rm s}}$. Only at two of these sites, $\ket{g_{\rm s}}$ and $\ket{j_{\rm s}}$, have lower on-site energy.
In other words, there are two potential wells at the sites $\ket{g_{\rm s}}$ and $\ket{j_{\rm s}}$ on the hypercube
and the terms $\hat{s}_m^x$ provides tunneling between them. So, it is clear that
the physics in each subspace ${\mathcal H}_{v_{j_{\rm s}}}$ is essentially double-well tunneling
in a hypercube with the initial state located at one of the wells $\ket{j_{\rm s}}$.

The Hamming distance between two binary arrays is the number of bits where they differ. 
We define the Hamming distance between $|g_{\rm s}\rangle$ and $\ket{j_{\rm s}}$ as $l_j$, which ranges from  0 to $n_{\rm s}$.
The dynamics in the subspaces with identical Hamming distance $l_j$ is exactly the same.
For larger $l_j$, the evolution time from $|j_{\rm s}\rangle$ to $|g_{\rm s}\rangle$ is longer.

The system described by the Hamiltonian in Eq. (\ref{eq:Hj2}) can be visualized roughly as a double-well system in Fig. \ref{fig:decay}(a).
For this kind of system, the low energy Hilbert space is  spanned by two wave packets $\ket{\chi_g}$
and $\ket{\chi_j}$ localized near $\ket{g_{\rm s}}$ and $\ket{j_{\rm s}}$, respectively. This is verified by our numerical computation.
In our numerical computation, we expand the interaction strength  in the polynomial form
\begin{eqnarray}
\lambda_{n_{\rm s}} &=& \frac{\gamma_1}{n_{\rm s}}+\frac{\gamma_2}{n_{\rm s}^2}+\frac{\gamma_3}{n_{\rm s}^3}+\cdots \ .
\end{eqnarray}
We then diagonalize numerically the Hamiltonian of Eq. (\ref{eq:Hj2}). As we expect that the two lowest eigenstates are of the form,
$\left(\ket{\chi_g}+\ket{\chi_j}\right)/\sqrt{2}$ and $\left(\ket{\chi_g}-\ket{\chi_j}\right)/\sqrt{2}$ if $j\neq g$, we superpose
them and obtain $\ket{\chi_g}$. As shown in  Fig. \ref{fig:decay}(c), we find that $\ket{\chi_g}$ is indeed localized and its
localization will not decrease as $n_{\rm s}$ increase if $\gamma_1\lesssim 1$ and $\gamma_2\lesssim 1.16$ .

The wave packet $\ket{\chi_g}$ can also be approximated analytically.
We rearrange the basis and write $\ket{\chi_g}$ as
\begin{eqnarray}
\ket{\chi_g}=\sum_{h=0}^{n_{\rm s}}\sum_{m=1}^{C_{n_{\rm s}}^h}a_{h,m}\ket{\psi_{h,m}},
\label{eq:xij}
\end{eqnarray}
where $|\psi_{h,m}\rangle$'s are re-arranged $\ket{i_{\rm s}}$'s with Hamming distance $h$ from
$|g_{\rm s}\rangle$, so that $|\psi_{h=0}\rangle=|g_{\rm s}\rangle$.
$m$ labels the different states with the same $h$. The $2^{n_{\rm s}}$ vertices of the hypercube
can be viewed as points on the surface of an $n_{\rm s}$-dimensional hypersphere
as seen in Fig. \ref{fig:decay}(b). There are $C_{n_{\rm s}}^h$ points  locating on the same latitude of the hypersphere,
which have the same $h$.
When $\ket{\chi_{j}}$ is far from $\ket{\chi_g}$ with $l_j\gg1$, the influence of $\ket{\chi_{j}}$ is so small that
 $\ket{\chi_g}$ has $n_{\rm s}$-fold rotation symmetry with the coefficients independent of $m$, i.e.,
\begin{eqnarray}
a_{h,m}\approx a_h\ .
\label{eq:ah}
\end{eqnarray}
Numerically computed $a_{h,m}$ are shown in Fig. \ref{fig:decay}(c), where  each blue point represents one $a_{h,m}$.
It is clear from the figure that the $C_{n_{\rm s}}^h$ points with the same $h$
are nearly indentical. They become visibly different  only near the location of $|j_{\rm s}\rangle$, i.e., at $h=9$  in
this example.
Most $a_{h,m}$ have the same sign except some near $|j_{\rm s}\rangle$.
The interaction $\sum_{m=0}^{n_{\rm s}-1}\hat{s}_m^x$ only changes one qubit, so each point at the $h$th will interact
with $h$ points at the $(h-1)$th and $n_{\rm s}-h$ points at $(h+1)$th as the yellow line shown in Fig. \ref{fig:decay}(b).
If we neglect the term $-\ket{j_{\rm s}}\bra{j_{\rm s}}$ using a tight-binding approximation, the eigen-equation
for Eq. (\ref{eq:Hj2}) can be written as
\begin{eqnarray}
-h\lambda_{n_{\rm s}} a_{h-1}+Va_h-(n_{\rm s}-h)\lambda_{n_{\rm s}} a_{h+1} = Ea_h \ ,
\label{eq:Hh}
\end{eqnarray}
where $V=-1$ if $h=0$ and $V=0$ if $h\geqslant1$. $a_h$ could be approached analytically using the iteration method [see Appendix \ref{app:wf}].

\begin{figure}[tb]
\begin{center}
\includegraphics[clip = true, width =0.9\columnwidth]{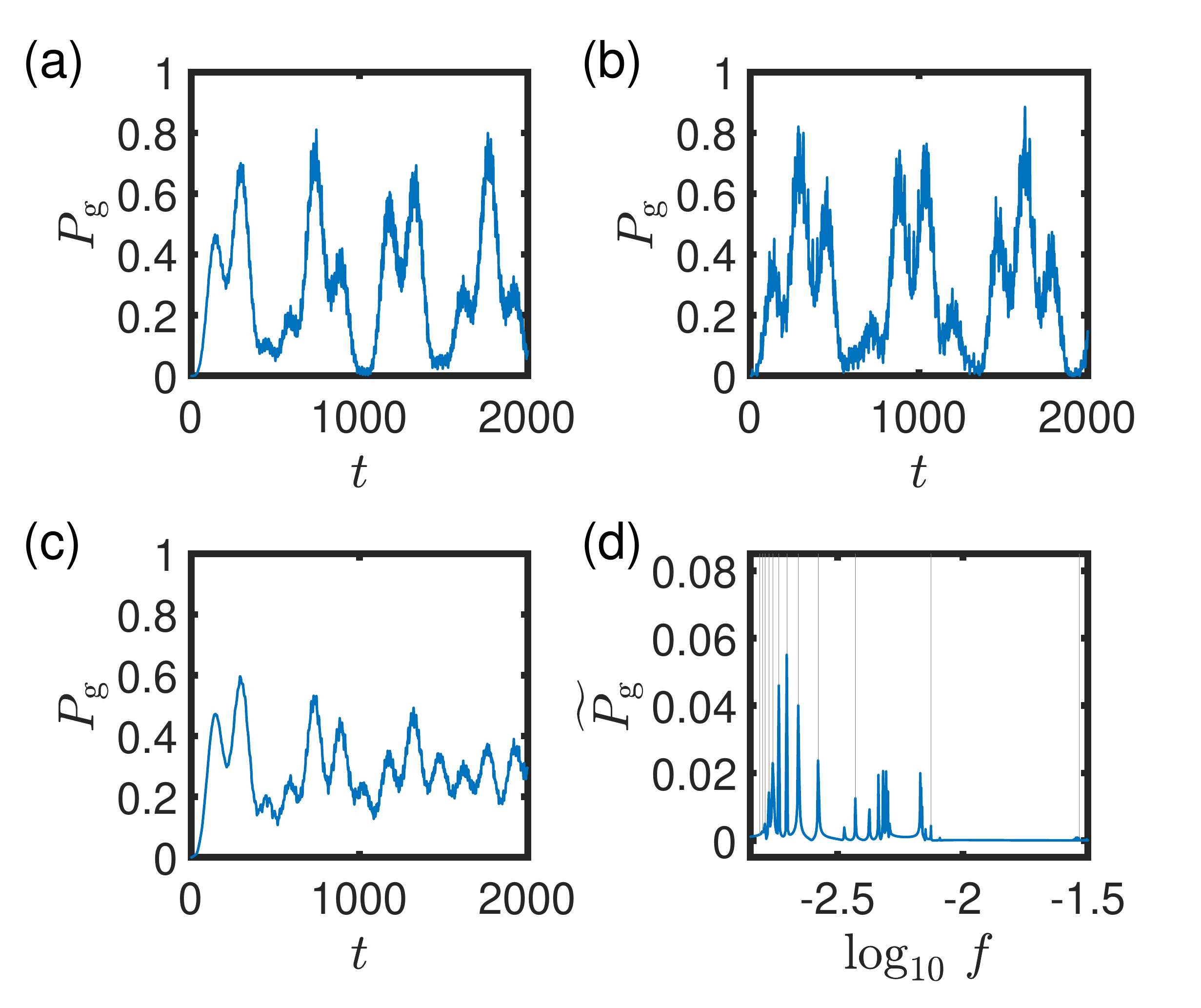}
\caption{\label{fig:osc} (color online) The evolution with Hamiltonian in Eq. (\ref{eq:H0}) and Eq. (\ref{eq:Hil}) with $n_{\rm s}=n_{\rm b}=12$. $\gamma_m$ is the same as Fig. \ref{fig:decay}. $P_g$ is the time dependent ground state probability of the problem system. For (a) and (b), the initial condition is $\ket{\Psi_{\rm in}}=\ket{j_{\rm s},g_{\rm b}}$ with Hamming distance (a) $l_j=6$, (b) $l_j=12$. (c) The initial state is $\ket{\Psi_{\rm in}}=\sqrt{1/N_{\rm s}}\sum_{j_{\rm s}=0}^{N_{\rm s}-1}\ket{j_{\rm s},g_{\rm b}}$. (d) The Fourier transformation of (c). The gray line point out the peaks contributed by different $l_j$. }
\end{center}
\end{figure}

The two wave packets $\ket{\chi_g}$ and $\ket{\chi_j}$ have the same on-site energy. Their interaction strength decides
the oscillation frequency $\omega_{l_j}=\left|\bra{\chi_g}H \ket{\chi_j}\right|$. In other words, $\omega_{l_j}$ is the
evolution speed from $|j_{\rm s}\rangle$ to $|g_{\rm s}\rangle$. Physically, the interaction should decay with Hamming
distance, i.e., $\omega_{l_j+1}<\omega_{l_j}$.

When  the problem system evolves into $| g_{\rm s}\rangle$ through tunneling from the initial state of Eq. (\ref{eq:sortin}),
it is cooled down by the bath and our goal is achieved.
It is clear that the larger the Hamming distance $l_j$ the longer it takes to get $| g_{\rm s}\rangle$. The longest time occurs when $l_j=n_{\rm s}$.
However, to have a detectable ground state probability, we just need to wait until half of the states with $l_j\leqslant\left\lfloor n_{\rm s}/2 \right\rfloor$
evolve to $| g_{\rm s}\rangle$. The ground state probability can  thus be approximated as
\begin{eqnarray}
P_g &\approx& \frac{1}{N_{\rm s}}\left(A_0+\sum_{l=1}^{\left\lfloor \frac{n_{\rm s}}{2} \right\rfloor} C_{n_{\rm s}}^lA_l\sin^2\omega_l t  \right) ,
\end{eqnarray}
where $t$ is in the time scale regime $1/\omega_{\left\lfloor n_{\rm s}/2 \right\rfloor}<t<1/\omega_{\left\lfloor n_{\rm s}/2  \right\rfloor+1}$
and $A_l$ is the oscillation amplitude of a scale around 1. On average, the ground state probability is
\begin{eqnarray}
\overline{P_g} &\approx& \frac{\overline{A_l}}{N_{\rm s}}\left(1+\sum_{l=1}^{\left\lfloor \frac{n_{\rm s}}{2} \right\rfloor} C_{n_{\rm s}}^l\overline{\sin^2\omega_l t} \right)\approx \frac{\overline{A_l}}{4}\,,
\end{eqnarray}
which is large enough for detection and independent of $n_{\rm s}$.


Fig. \ref{fig:osc}(a) displays the oscillations of ground state probability with $l_j=\left\lfloor n_{\rm s}/2 \right\rfloor$ and
Fig. \ref{fig:osc}(b) shows the oscillations with $l_j=n_{\rm s}$. The period of (b) is larger than (a) because of longer Hamming distance. The oscillations with $l_j=n_{\rm s}$ has largest time scale which corresponds to the full thermal equilibrium. The evolution with the initial state Eq. (\ref{eq:sortin}) is shown in Fig. \ref{fig:osc}(c), where
the increasing slope near $t=0$ is seen similar to (a). It indicates that the problem system can be cooled down considerably
earlier before the equilibrium between the bath and problem system is reached. Fig. \ref{fig:osc}(d) is the Fourier transformation of (c).
You can clearly see the peaks for independent oscillations with different $l_j$. 

\begin{figure}[tb]
\begin{center}
\includegraphics[clip = true, width =0.8\columnwidth]{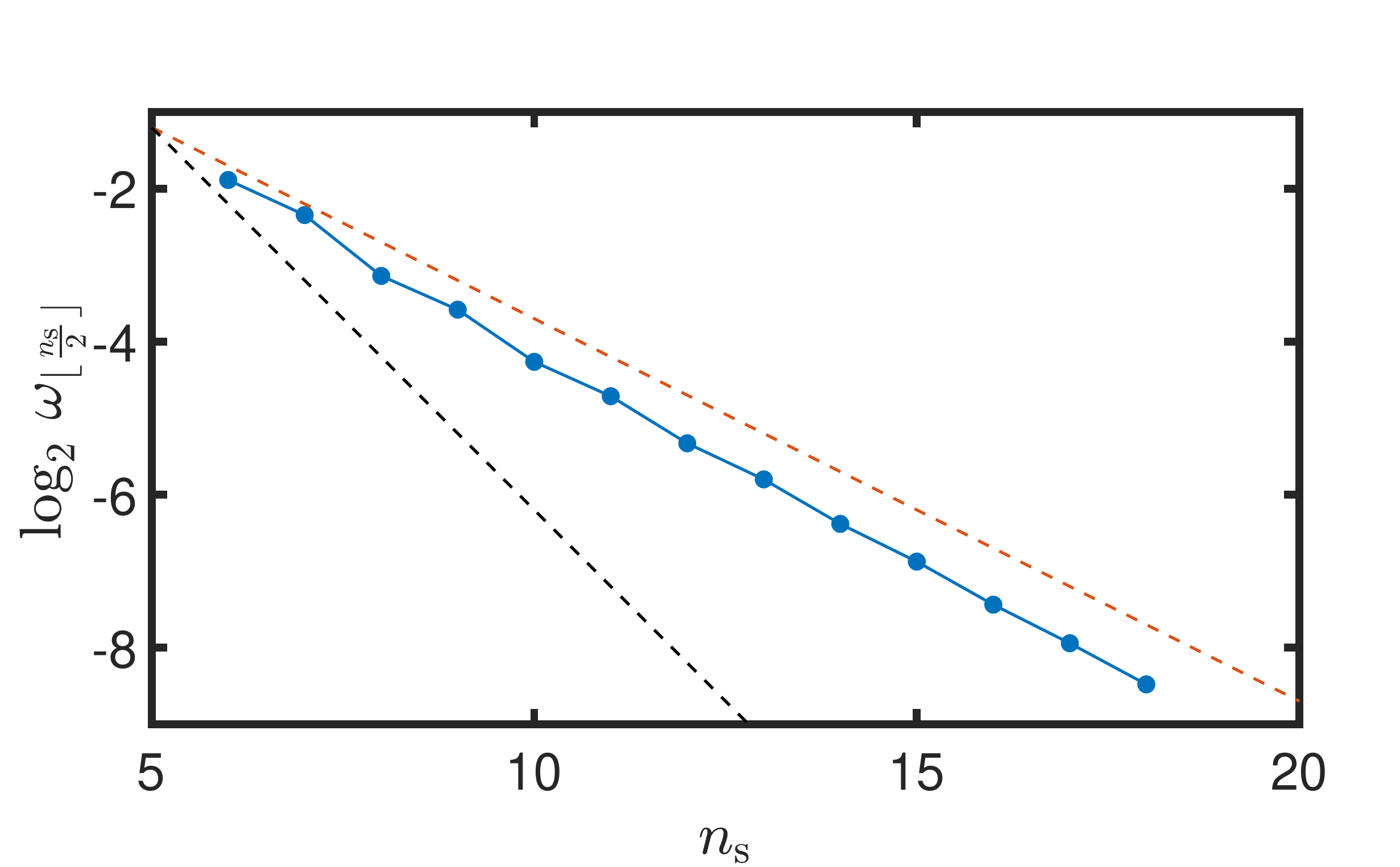}
\caption{\label{fig:w} (color online) The oscillation  frequency between $|j_{\rm s}\rangle$ and $|g_{\rm s}\rangle$
with distance $l_j=\left\lfloor n_{\rm s}/2\right\rfloor$ according to Eq. (\ref{eq:Hj2}). The blue dots are the exact
value from diagonalization whose slop is about -0.55. $\gamma_m$ is the same as Fig. \ref{fig:decay}.
The slope of the orange dash line is -0.5 representing Grover's algorithm. The slope of the black dash line is -1 representing the classical algorithm.}
\end{center}
\end{figure}

The oscillation  frequency $\omega_{l_j}$ is decided by the energy difference of two lowest energy states in the subspace.
The cooling speed is about $\omega_{\left\lfloor n_{\rm s}/2  \right\rfloor}$. Fig. \ref{fig:w} shows the
dependence of the cooling speed on the number of qubits $n_{\rm s}$.
It is calculated by numerical diagonalizing Eq. (\ref{eq:Hj2}) with $l_j=\left\lfloor n_{\rm s}/2 \right\rfloor$.
By fitting the numerical result, we find that the cooling speed is about $O(N_{\rm s}^{0.55})$ with local interaction, which is close to
the Grover's algorithm~\cite{Grover1996}. The form of interaction does not strongly affect the cooling speed.

\section{Discussion and Conclusion}
We have proposed a general framework of quantum computing by cooling a Hamiltonian system whose
ground states encode the solutions of a given problem with a fully quantum (non-Markovian) bath. 
This bath, which could be called a quantum icebox,  is an interacting spin system with trivial and 
easy-to-prepare ground states so that it  can be brought close to absolute zero temperature readily. 
We illustrated this method in two specific realizations in the benchmark problem of unsorted search.  
In both cases, we found a strong quantum advantage.

It is appropriate to contrast our work with the more familiar quantum adiabatic algorithm (QAA) or quantum annealing \cite{Johnson2011,McGeoch2014,QiuXingze2020}.
In QAA, the system with simple Hamiltonian is set to its simple ground state (effectively absolute zero temperature), and it is 
then slowly changed (or annealed) to another more complicated Hamiltonian, whose ground states are the solutions
of a given problem \cite{Farhi2001,Lucas2014}. In the whole process, the system is vulnerable to external heat or noise \cite{Paladino2014,Bilmes2017,Braumuller2020} and often encounters 
exponentially small energy gap \cite{Young2008}. In our framework, the quantum icebox can be made large enough to offer two advantages:
(1) to make sure the quantum icebox does not heat up before the system cools down; (2) to protect the system from decoherence. 
Given that a functioning quantum computer has already been built based QAA \cite{Harris2010,Johnson2011}, our icebox strategy seems likely to be practicable. Specifically, the experimental systems used to implement 
QAA and quantum simulation (QS) \cite{Monroe2021,Ebadi2021} can be modified 
to explore this possibility. 

Heat transfer has long been regarded as a stochastic thermal process \cite{HeXiaoyi1998,Wang2008,Saaskilahti2013}. Our quantum icebox shows that cooling can 
be done coherently. It raises fresh questions about the connection between heat transport and the flow of  quantum information. 

\acknowledgments
 FW is supported in part by the U.S. Department of Energy under grant DE-SC0012567, by the European Research Council
 under grant 742104, and by the Swedish Research Council under contract 335-2014-7424.  BW and JF are supported by the National Key R\&D Program of China (Grants No.~2017YFA0303302, No.~2018YFA0305602), National Natural Science Foundation of China (Grant No. 11921005), and Shanghai Municipal Science and Technology Major Project (Grant No.2019SHZDZX01).

~\\

\appendix

\section{Analytical result of spin wave propagation \label{app:wave}}
The spin wave dynamics with total Hamiltonian $H=H_{\rm s}+H_{\rm b}+H_{\rm I}$ of Eq. (\ref{eq:XXX}) and (\ref{eq:Hone})
is illustrated numerically in Fig. \ref{fig:wave}. It can also be demonstrated  in the single  excited mode approximation, where
 we consider just states $\ket{e_{\rm s}, k_g}$ and $\ket{g_{\rm s}, k}$, neglecting the states with multi-magnon.
 $\ket{g_{\rm s}}$ and $\ket{e_{\rm s}}$  represents the excited state and ground state of the problem system, $\ket{k_g}$ and $\ket{k}$
 represent the ground state and the excited states of the bath with wave vector $k$.
The Hamiltonian becomes
\begin{eqnarray}
H &=& \sum_{k\neq k_g}\big[\Delta E_k \ket{g_{\rm s}, k}\bra{g_{\rm s}, k} \nonumber\\
&&+\lambda_{k}\ket{g_{\rm s}, k}\bra{e_{\rm s}, k_g}+\lambda_{k}^*\ket{e_{\rm s}, k_g}\bra{g_{\rm s}, k} \big] \ ,
\end{eqnarray}
where $\Delta E_k=\left( E_{ {\rm s}g }+E_{ {\rm b}k } \right)-\left( E_{ {\rm s}e }+E_{ {\rm b}k_g } \right)$ is the energy detuning and $\lambda_k$ is the coupling strength.
The time dependent wave function is
\begin{eqnarray}
\ket{\Psi} &=& b_g \ket{e_{\rm s}, k_g}+\sum_{k\neq k_g}b_k\ket{g_{\rm s}, k} \ ,
\end{eqnarray}
where the probability amplitude satisfies the Schr\"{o}dinger equation with
\begin{eqnarray}
\left\{
\begin{array}{ccc}
  i\frac{{\rm d} b_g}{{\rm d} t}&=&\sum_{k\neq k_g}\lambda_k^*b_k \\
  i\frac{{\rm d} b_k}{{\rm d} t}&=&\Delta E_k b_k+\lambda_kb_g
\end{array}\right.  \ .
\label{eq:Schb}
\end{eqnarray}
In the early time of evolution $|\lambda_{k}|t\rightarrow0$, $b_g\approx1$. We can decouple the equations and get~\cite{Zhang2016}
\begin{eqnarray}
  b_k &\approx& \frac{\lambda_k}{\Delta E_k}\left( e^{-i\Delta E_kt}-1 \right) \ .
\end{eqnarray}
The probability amplitude for the two level system in its excited state is
\begin{eqnarray}
 b_g&\approx&1-2\sum_{k\neq k_g}\frac{|\lambda_k|^2}{\Delta E_k^2}\sin^2\frac{\Delta E_kt}{2}
\end{eqnarray}

In the position coordinate, the wave function is
\begin{eqnarray}
\phi &\approx& b_ge^{ik_gx}+\sum_{k\neq k_g}b_k e^{ikx} \ .
\label{eq:phix}
\end{eqnarray}
The phase difference between different $b_k$ changes with time, the wave function will spread out from $x=0$.

\section{Non-locality of  Hamiltonians\label{app:oracle}}
The Hamiltonians used in quantum algorithms must be physically reasonable. This usually means that the Hamiltonian are
$k$-local, i.e.,  contain only interactions involving no more than a fixed number $k$ of qubits~\cite{Dam2007}.
Although the three Hamiltonians in Eqs. (\ref{eq:H0},\ref{eq:Hinl}) are not $k$-local,
they are physically reasonable, and here is the explanation.

In the Grover's algorithm, a single Grover iteration is $U_{\rm G}=R_\xi R_g$ \cite{Nielsen2010}. $R_g=\mathbb{I}-2\ket{g}\bra{g}$ is
the oracle operator for the target $\ket{g}$. And
$R_\xi=H_{\rm a}^{\otimes n}\left(\mathbb{I}-2\ket{0}\bra{0}\right)H_{\rm a}^{\otimes n}=\mathbb{I}-2\ket{\xi}\bra{\xi}$,
where $H_{\rm a}$ is the Hadamard gate and $\ket{\xi}=\sqrt{1/N}\sum_{j=0}^{N-1}\ket{j}$.

That the Hamiltonians in Eqs. (\ref{eq:H0},\ref{eq:Hinl}) are reasonable despite being non-local is because the dynamics generated by them
can be implemented with the Grover operation $U_{\rm G}$. For simplicity, we consider the Hamiltonian dynamics $U=e^{-iHt}$
with $H=-\ket{g}\bra{g}-\ket{\xi}\bra{\xi}$. When the time evolution is  discretized with  time step  $\Delta t=\pi$ and $T=m\Delta t$,
we have \cite{Mochon2007}
\begin{eqnarray}
U=e^{-iHT}  \approx \Pi_{j=1}^m e^{-i\pi H} \approx \Pi_{j=1}^m U_{\rm G} \ .
\end{eqnarray}
Note that the circuit complexity for implementing the oracle is $O\left(n^3 \right)$ \cite{TANAKA2011} and
the time complexity is $O\left(n^2\right)$ \cite{Ito2014}, where $n = \log_2 N$.  \\

\section{Exact Hamiltonian for the non-local model\label{app:H44}}
We expand the total Hamiltonian $H=H_{\rm s}+H_{\rm b}+H_{\rm I}$ of Eq. (\ref{eq:H0}) and (\ref{eq:Hinl}) in terms of
 $\ket{Y}$, $\ket{\beta}$, $\ket{\alpha}$, $\ket{G}$ of Eqs. (\ref{eq:Y}), (\ref{eq:beta}), (\ref{eq:alpha}) and (\ref{eq:G}). Its exact matrix is
\begin{widetext}
\begin{eqnarray}
H=-\frac{1}{N_{\rm c}}\left(
\begin{smallmatrix}
  N_{\rm c}-N_{\rm s}-N_{\rm b}+1 & \sqrt{(N_{\rm c}-N_{\rm s}-N_{\rm b}+1)(N_{\rm b}-1)} & \sqrt{(N_{\rm c}-N_{\rm s}-N_{\rm b}+1)(N_{\rm s}-1)} & \sqrt{N_{\rm c}-N_{\rm s}-N_{\rm b}+1} \\
  \sqrt{(N_{\rm c}-N_{\rm s}-N_{\rm b}+1)(N_{\rm b}-1)} & N_{\rm c}+N_{\rm b}-1 & \sqrt{(N_{\rm s}-1)(N_{\rm b}-1)} & \sqrt{N_{\rm b}-1} \\
  \sqrt{(N_{\rm c}-N_{\rm s}-N_{\rm b}+1)(N_{\rm s}-1)} & \sqrt{(N_{\rm s}-1)(N_{\rm b}-1)} & N_{\rm c}+N_{\rm s}-1 & \sqrt{N_{\rm s}-1} \\
  \sqrt{N_{\rm c}-N_{\rm s}-N_{\rm b}+1} & \sqrt{N_{\rm b}-1} & \sqrt{N_{\rm s}-1} & 2N_{\rm c}+1
\end{smallmatrix} \right) \ .
\end{eqnarray}
\end{widetext}
If we just keep the leading terms in the limit of $1\ll N_{\rm s}, N_{\rm s}  \ll N_{\rm c}$, it recovers  Eq. (\ref{eq:simH44}) in the main text. Its  eigen-energies are
\begin{eqnarray}
E_3&=&-1+\sqrt{\frac{N_{\rm s}+N_{\rm b}}{N_{\rm c}}} \ ,  \\
E_2&=&-1 \ ,  \\
E_1&=&-1-\sqrt{\frac{N_{\rm s}+N_{\rm b}}{N_{\rm c}}} \ ,  \\
E_0&=&-2 \ .
\end{eqnarray}
The eigen-states are
\begin{eqnarray}
\ket{\Psi_0}&=&
\begin{pmatrix}[1.8]
  0 \\
  0 \\
  0 \\
  1
\end{pmatrix} \,, ~
\ket{\Psi_1}=
\begin{pmatrix}[1.8]
  \frac{1}{\sqrt{2}} \\
  \sqrt{\frac{N_{\rm s}}{2(N_{\rm s}+N_{\rm b})}} \\
  \sqrt{\frac{N_{\rm b}}{2(N_{\rm s}+N_{\rm b})}} \\
  0
\end{pmatrix} \ ,
\end{eqnarray}

\begin{eqnarray}
\ket{\Psi_2}&=&
\begin{pmatrix}[1.8]
  0 \\
  -\sqrt{\frac{N_{\rm b}}{N_{\rm s}+N_{\rm b}}} \\
  \sqrt{\frac{N_{\rm s}}{N_{\rm s}+N_{\rm b}}} \\
  0
\end{pmatrix} \ ,~
\ket{\Psi_3}=
\begin{pmatrix}[1.8]
  -\frac{1}{\sqrt{2}} \\
  \sqrt{\frac{N_{\rm s}}{2(N_{\rm s}+N_{\rm b})}} \\
  \sqrt{\frac{N_{\rm b}}{2(N_{\rm s}+N_{\rm b})}} \\
  0
\end{pmatrix}\ .
\end{eqnarray}
The time dependent wave function with initial condition $\ket{\Psi_{\rm in}}=\sqrt{1/N_{\rm s}}\ket{G}+\sqrt{(N_{\rm s}-1)/N_{\rm s}}\ket{\alpha}$ is
\begin{eqnarray}
&&\ket{\Psi}=e^{-iE_0 t}\sqrt{\frac{1}{N_{\rm s}}}\ket{\Psi_0}+ \nonumber\\
&&\sqrt{\frac{N_{\rm s}-1}{N_{\rm s}}} \frac{\sqrt{\frac{N_{\rm s}}{2}} \left( e^{-iE_3 t}\ket{\Psi_3}+e^{-iE_1 t}\ket{\Psi_1} \right)-e^{-iE_2 t}\sqrt{N_{\rm b}}\ket{\Psi_2}
}{\sqrt{N_{\rm s}+N_{\rm b}}} \ . \nonumber\\
\end{eqnarray}
Expanding it, we will get Eq. (\ref{eq:Psit}) in the main text.

\section{Exclusive-OR (XOR)\label{app:xor}}
The module 2 addition "$\oplus$" is also called XOR operation  for two Boolean variables, which is defined as $0\oplus0=1\oplus1=0$ and $0\oplus1=1\oplus0=1$.
For an integer $i$, its binary digits $i^{(m)}$'s are defined as
\begin{eqnarray}
i =\sum_{m=0}^{n-1}2^mi^{(m)}\ .
\end{eqnarray}
For any two integers $i_a$ and $j_b$, $v=i_a\oplus j_b$ is defined bitwise as,
\begin{equation}
v^{(m)}=i_a^{(m)}\oplus j_b^{(m)}\ .
\end{equation}
It also can be written as
\begin{eqnarray}
v = i_a\oplus j_b=\sum_{m=0}^{n-1}2^m\left(i_a^{(m)}\oplus j_b^{(m)}\right) \  .
\end{eqnarray}
For example $12\oplus10=1100\oplus1010=0110=6$. There is an inverse relation that $j_b=i_a\oplus v$ if $v=i_a\oplus j_b$.
We can check that $12\oplus6=1100\oplus0110=1010=10$.

\section{\label{app:wf} Analytic approximation of the wave packet $\ket{\chi_g}$}
 Eq. (\ref{eq:Hh}) can be approached analytically. We re-write it as
\begin{eqnarray}
\left\{
\begin{array}{l}
(-1-E)a_0-n_{\rm s}\lambda a_1 = 0 \\
-h\lambda a_{h-1}-Ea_h-(n_{\rm s}-h)\lambda a_{h+1} = 0 \ (1\leqslant h \leqslant n_{\rm s} )
\end{array}\right. \ .
\end{eqnarray}
We define the ratio $b_h=a_h/a_{h-1}$ and get
\begin{eqnarray}
b_h = \frac{h\lambda}{1+n_{\rm s}\lambda b_1-(n_{\rm s}-h)\lambda b_{n+1}} \  .
\end{eqnarray}
With the self-consistent method,  the above iteration becomes
\begin{eqnarray}
b_h^{(m+1)} = \frac{h\lambda}{1+n_{\rm s}\lambda b_1^{(m)}-(n_{\rm s}-h)\lambda b_{h+1}^{(m)}} \ ,
\end{eqnarray}
where the superscript is the order of approximation. If we set $b_h^{(0)}=0$, we get
\begin{eqnarray}
    b_h^{(1)}&=& h\lambda \ , \\
   b_h^{(2)} &=& \frac{h\lambda}{1+n_{\rm s}\lambda^2-(n_{\rm s}-h)(h+1)\lambda^2 } \ , \\
    b_h^{(3)} &=&  \frac{h\lambda}{1+\frac{n_{\rm s}\lambda^2}{1-(n_{\rm s}-2)\lambda^2 } -\frac{(n_{\rm s}-h)(h+1)\lambda^2}{1+n_{\rm s}\lambda^2-(n_{\rm s}-h-1)(h+2)\lambda^2} }\ .
\end{eqnarray}
The corresponding energy is
\begin{eqnarray}
    E^{(1)}&=& -1-n_{\rm s}\lambda^2 \ , \\
   E^{(2)} &=& -1-\frac{n_{\rm s}\lambda^2}{1-(n_{\rm s}-2)\lambda^2 } \ , \\
    E^{(3)} &=& -1- \frac{n_{\rm s}\lambda^2}{1+\frac{n_{\rm s}\lambda^2}{1-(n_{\rm s}-2)\lambda^2  } -\frac{2(n_{\rm s}-1)\lambda^2}{1-(2n_{\rm s}-6)\lambda^2} } \ .
\end{eqnarray}

We define $a_0=1/\sqrt{\aleph}$, where $\aleph$ is the normalization factor. The coefficient for $h\geqslant1$ is
\begin{eqnarray}\label{eq:aexact}
a_h&=&\frac{1}{\sqrt{\aleph}}\prod_{m=1}^h b_m \ .
\end{eqnarray}
The first order approximation is
\begin{equation}\label{eq:a1exact}
a_h^{(1)}=\frac{1}{\sqrt{\aleph^{(1)}}} \frac{h!}{n_{\rm s}^h}\approx\sqrt{\frac{2\pi h}{\aleph^{(1)}}}\left( \frac{ h}{en_{\rm s}}\right)^h\,,
\end{equation}
where the last term is obtained with the Stirling approximation.
It is accurate only in the regime  $h\geqslant1$ and $h/n_{\rm s}<0.2$.
By fitting the numerical data in Fig. \ref{fig:decay}(c), we find that
the decay speed of $a_h$ is exponential  in the regime $h/n_{\rm s}<0.2$ and
inversely proportional to $h$ in the regime $h/n_{\rm s}>0.3$.

\normalem




\end{document}